\documentclass[12pt]{article} 
\usepackage{graphicx} 
\usepackage{times}
\usepackage{caption} 
\usepackage{siunitx}
\usepackage{subcaption} 
\usepackage{bm}
\usepackage{color,soul}
\usepackage[super,numbers,sort&compress,elide]{natbib}

\newenvironment{sciabstract}{%
\begin{quote} \bf}
{\end{quote}}

\newcommand{\ket}[1]{\ensuremath{|{#1}\rangle}}
\newcommand{\bra}[1]{\ensuremath{\langle{#1}|}} 
\newcommand{\braket}[2]{\ensuremath{\langle{#1}\vert{#2}\rangle}}     \newcommand{\eqn}[1]{\begin{equation}{#1}\end{equation}}

\newcommand{\iso}[2]{\textit{{#1}}$\rightarrow$\textit{{#2}}}

\title{Strong coupling with light enhances the photoisomerization quantum yield of azobenzene} 
\author{Jacopo Fregoni,$^{1,2}$ Giovanni Granucci,$^{3\ast}$ Maurizio Persico$^{3}$ and Stefano Corni$^{2,4,5,\ast\ast}$\\
\\
\footnotesize{$^{1}$ Dipartimento di Scienze Fisiche, Informatiche e Matematiche, Universit\`a di Modena and Reggio Emilia,}\\
\footnotesize{I-41125 Modena, Italy}\\
\footnotesize{$^{2}$ Istituto Nanoscienze, Consiglio Nazionale delle Ricerche CNR-NANO, I-41125 Modena, Italy}\\
\footnotesize{$^{3}$ Dipartimento di Chimica e Chimica Industriale, Universit\`{a} di Pisa, I-56124 Pisa, Italy}\\
\footnotesize{$^{4}$ Dipartimento di Scienze Chimiche, Universit\`{a} di Padova, I-35131 Padova, Italy}\\
\footnotesize{$^{5}$ Lead Contact}\\
\footnotesize{$^\ast$ Correspondence: giovanni.granucci@unipd.it}\\
\footnotesize{$^{\ast\ast}$ Correspondence: stefano.corni@unipd.it}}

\date{}
\begin{document} \baselineskip24pt 
\maketitle 
\begin{sciabstract} 

\section*{Summary}

The strong coupling between molecules and photons in resonant cavities offers a
new toolbox to manipulate photochemical reactions. While the quenching of
photochemical reactions in the strong coupling regime has been demonstrated before, their enhancement has proven to be more elusive. By means of a
state-of-the-art approach, here we show how the \iso{trans}{cis} photoisomerization quantum yield of azobenzene embedded in a realistic environment can be higher in polaritonic conditions than in the cavity-free case. We characterize the
mechanism leading to such enhancement and discuss the conditions to push the
photostationary state towards the unfavoured reaction product. Our results
provide a signature that the control of
photochemical reactions through strong coupling can be extended from selective
quenching to improvement of the quantum yields.
\end{sciabstract}

\section*{Introduction}

The interaction between light and matter at the nanoscale is at the basis of a
manifold of experimental applications in
plasmonics\citep{fang:waveguides,dahlin:nanoplas,stockman:ultrafastnanoplas,baumberg:tunneling},
single-molecule spectroscopies\citep{stockman:nanomicroscop,vanhulst:antenna},
nanoprinting\citep{wegener:nanoprint} and
nanocavity optics\citep{vanhulst:singleemitter,vanhulst:cavity,vanhulst:nanocavity}.
When light is sufficiently confined in micro/nanometric systems in presence of one or more
quantum emitters, its exchange of energy with the emitters becomes
coherent and the system enters the strong coupling
regime\citep{JC:cav,flick:weaktostrong}. Accordingly, the degrees of freedom of
light and matter mix and the states of the system are described as hybrids
between the two: the polaritons\citep{feist:modific,aizpurua:coherentcoupling}.
The first experimental realizations to pioneer the idea of controlling the chemical reactions through strong coupling of molecules with light made use of metallic cavities\citep{ebbesen:switch,ebbesen:molecsc}. Later on, the achievement of strong coupling with plasmonic nanocavities at the single-molecule level at room temperature has been obtained with a Nanoparticle on a Mirror (NPoM) setup\citep{baumberg:singlemol,baumberg:quenching}. Such setup has been recently improved with DNA origami for higher reproducibility \citep{baumberg:dnaorigami,baumberg:qed}. The manifold of possibilities opened up by such experiments drove efforts to explore microcavities-based setups at low temperature, achieving longer lifetimes for the whole system\citep{wang:singlemicrocav}. Theoretical modeling followed immediately to survey the plethora of new possibilities offered by strong light-molecule coupling\citep{flick:review, ribeiro:polaritonchem}. 
The high flexibility of the polaritonic properties has been assessed for both realized\citep{ebbesen:workfun,stellacci:ultrastrong,baumberg:qed} 
and potential applications\citep{feist:modific,aizpurua:singleplas} giving birth to a new branch of chemistry\citep{mukamel:novel}: the so-called polaritonic chemistry.\citep{feist:polchem}\\

When a resonant mode is coupled to electronic transitions, the molecules
exhibit enhanced spontaneous emission at both the collective and single
molecule level\citep{strauf:lasing,sawant:lasing,du:lasing,gies:emission}. When the coupling is sufficiently strong, coherent energy exchange occurs between light and photoactive molecules, potentially translating into
modified photochemical properties\citep{ebbesen:switch}. The
modifications to the potential energy surfaces (PESs through all the current work) driving different photophysical
and photochemical behaviours are described by a basis of direct products of electronic and photonic states. Under this
assumption, the states of the system are best described as hybrids between electronic and
photonic.\citep{flick:weaktostrong,aizpurua:coherentcoupling,deliberato:ultrastrong}.\\

\begin{figure*}[h!]{\includegraphics[width=1\linewidth]{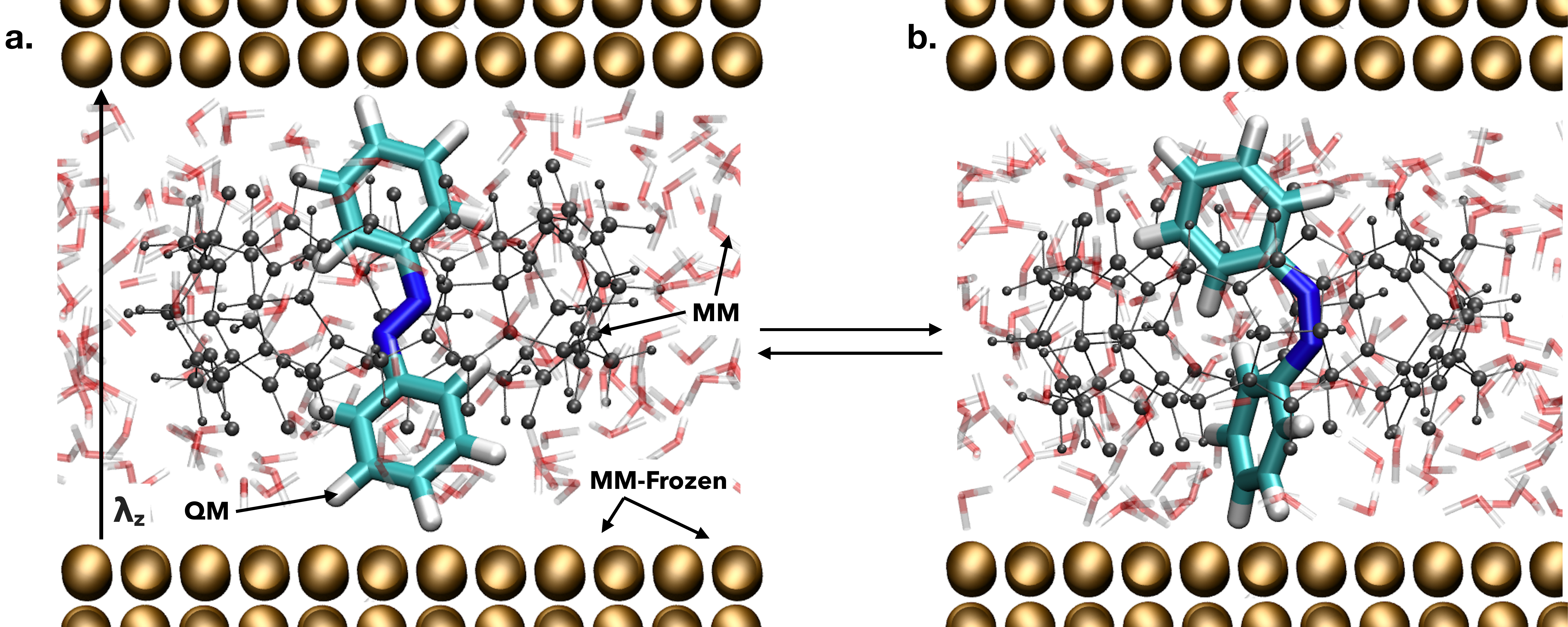}}
	\caption{\textbf{Simulated system} Snapshots of the simulated system mimicking a plasmonic nanocavity as the one reported by Baumberg and coworkers\citep{baumberg:singlemol}. The molecule, here azobenzene in \textit{trans} (panel \textbf{a}) or \textit{cis} (panel \textbf{b}) configurations, is computed at QM level (see text and Methods) and interacts with the MM environment by electrostatic embedding plus Lennard-Jones potentials. The environment is composed by cucurbit-7-uril (grey organic molecule cage) and gold layers (four layers on each side, frozen MM degrees of freedom), including also explicit water molecules. The cavity mode is polarized along $\rm{\bm{\lambda_z}}$ and the sampling is run at room temperature\citep{bussi:thermo}.}\label{fig:env} 
\end{figure*}

The possibility to shape the electronic states with quantum light inspired various groups to explore the role of strong light-molecule coupling in controlling photochemical processes. For collective effects, the focus has been on polariton formation in full quantum diatomic molecules\citep{vendrell:mctdhsc} and on several model dye molecules in a realistic environment\citep{luk:multiscale}. At the single molecule level, the non-adiabatic dynamics schemes developed allowed to predict features arising on the PESs like the creation of
avoided crossings and light-induced conical intersections\citep{feist:modific,
vibok:lightcoin,vibok:lightcoin2}. Such features modify the shape of PESs, translating into a potentially different photochemical reactivity\citep{spano:coherent,mukamel:novel,mukamel:nadiab,galego:suppressing}. The possibility to enhance the yield of photochemical processes has been recently proven for energy transfer\citep{spano:coherent,feist:et}, singlet fission\citep{ribeiro:singletsplit} and catalysed reactions through vibrational strong coupling, obtained by exploiting remote catalysts\citep{du:remote}. For strong coupling with resonant optical frequencies, enhancement has only been suggested by calculations on model PESs\citep{feist:polchem,feist:manymols} and neglecting the cavity losses and realistic non-radiative events.\\ 
s
As such events play a central role in the yields of photochemical reactions, the question remains if strong coupling can lead to a real enhancement of photochemical quantum yields in real molecules.
 Even more practically, the interest resides in the photostationary regime and in determining whether the related
concentrations of products is enriched with respect to the standard reaction conditions. Here, by means of the state-of-the-art approach we devised\citep{fregoni:manipulating}, we show that it is possible to
identify conditions that lead to improved quantum yields and product-enriched
photostationary states. 
By investigating azobenzene
\iso{trans}{cis} photoisomerization in strong coupling, we compare to the zero coupling case and highlight the differences between the two processes.
Such comparison allows us to propose an interpretation of the mechanism leading
to the increased quantum yield for the \iso{trans}{cis} $\pi-\pi^*$
photoisomerization.\\ 

The model system we simulate is depicted in Figure \ref{fig:env} and mimics the experimental setup used by Baumberg and coworkers\citep{baumberg:singlemol} for achieving strong coupling with a single methylene blue chromophore. The azobenzene molecules are hosted in a one-to-one arrangement by cucurbit-7-uril ring molecules, which are in turn adsorbed on a planar gold surface. In this arrangement, the azobenzene long axis is approximately perpendicular to the surface. This is relevant because the field polarization, and the transition dipole for the $S_0$-$S_1$ and the $S_0$-$S_2$ transitions are all aligned in the same direction\citep{persico:oscill}. The cavity is completed by gold nanoparticles sitting on top of the cucurbituril ring and much larger than the latter, so we simulate them as a second planar surface. Explicit water molecules fill the space between the gold layers (see Supplemental Note 1). 

\section*{Results}
\subsection*{Polaritons in {azobenzene}}

Before investigating the photochemical properties of molecules under strong
coupling, we show how the coupling conditions affect the energy landscape in
the case of multiple electronic states.
In this section, we aim only to provide an interpretative framework for the results of next section, and hence the results presented in this section are computed without environment.\\

\begin{figure*}[htt!]{\includegraphics[width=0.8\linewidth]{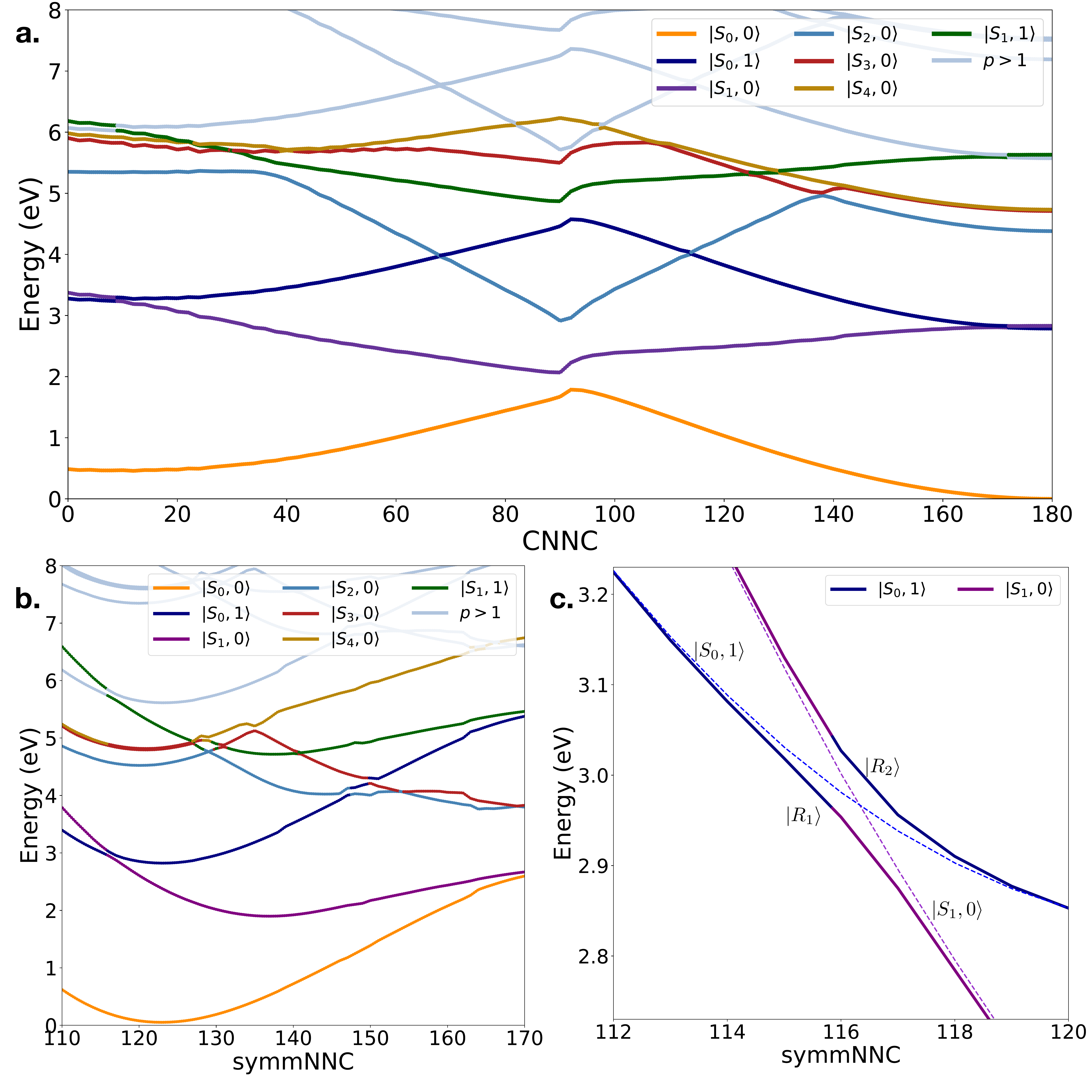}}
	\caption{\textbf{Polaritonic potential energy curves of azobenzene with photon of 2.8 eV and single-photon electric field strength $E_{1ph}$ of 0.002 a.u.} The polaritonic PESs are computed at the \textbf{a)} CNNC coordinate while relaxing all the other degrees of freedom for the ground state and \textbf{b)} symNNC coordinate with CNNC 165$^\circ$ and constrained optimization as before. Each polaritonic branch is colored depending on the uncoupled state that majorly composes the polariton at each geometry. \textbf{c)} Detail of the strong coupling avoided crossing along the symNNC coordinate and between the $\ket{S_0,1}$ and $\ket{S_1,0}$ states where this feature gives rise to a different mechanism for the photoisomerization.}\label{fig:mol_comp} 
\end{figure*}

In Figure \ref{fig:mol_comp} we present two relevant cuts of the polaritonic PESs for the isolated azobenzene molecule, one along the CNNC dihedral and the other along the symmetric NNC bending (symNNC). In the former, all other degrees of freedom and also symNNC were optimized for the ground state.  In the latter, the analogous constrained optimization was done for each symNNC value, except that CNNC was fixed at 165$^\circ$, in order to show a clear cut of the polaritonic avoided crossing modifying the dynamics (notice that at the trans planar geometry the $S_0$,$S_1$ transition dipole moment vanishes). We shall exploit the PESs presented in this section to act as a qualitative and conceptual aid. By doing so, we introduce the framework to discuss the mechanism leading to the enhanced yield of the photoisomerization reaction under realistic environment.\\

Even in absence of environment, when a single molecule is strongly coupled with a cavity, polaritons drastically affect its PESs\citep{flick:weaktostrong,feist:modific,fregoni:manipulating}. The
photochemical properties are, in turn, deeply affected by the shape of the
polaritonic PESs. Aiming to thoroughly describe the molecule in the strong
coupling regime, we build the polaritonic Hamiltonian in the framework of a semiempirical wavefunction method\citep{granucci:fomograd}:\\

\eqn{
	\hat{H}_{{tot}}=\hat{H}_{{mol}}+\hat{H}_{{cav}}+\hat{H}^{sc}_{{int}}.
\label{eq:hamiltonian}} 

Here $\hat{H}_{mol}$ is the semiempirical electronic Hamiltonian,
$\hat{H}_{{cav}}$ is the quantized electromagnetic field Hamiltonian for an effective resonant mode set at optical frequencies and $\hat{H}^{{sc}}_{{int}}$ is the quantum interaction between
light and molecule in a dipolar fashion:

\eqn{
	\hat{H}^{sc}_{{int}}=E_{1ph}\sum_{n\ne n'}\ket{{n}}\bm{\lambda}\cdot\bm{\mu}(\bm{R})_{{n,n'}}\bra{{n'}} \left( \hat{b}^\dagger + \hat{b} \right).\label{eq:inter}
}

$E_{1ph}$ represents the magnitude of the single-photon electric field of the confined light mode, $\bm{\mu}_{\rm{n,n'}}$ is the transition
dipole moment between the electronic states, $\bm{\lambda}$ is the field
polarization unit vector, $\hat b^\dagger$ and $\hat b$ are the bosonic creation and annihilation operators. The nuclear motion is treated classically, using the surface-hopping approach\citep{fregoni:manipulating} (see Methods). By relying on a semiempirical wavefunction method, we provide a
detailed description of the electronic structure at low computational cost.
Such electronic structure method exploits a solid parameterization
\citep{cusati:semiemp} of the semiempirical electronic Hamiltonian and has been
previously validated against experimental data in a number of applications\citep{cusati:photodynamics,cantatore:yields,favero:app,benassi:gold}.\\

To gain more insight on the polaritonic PESs features, we refer to the basis of uncoupled products of light and matter
wavefunctions, given by the diagonalization of $\hat H_{mol} + \hat H_{cav}$, labeled as $\ket{n,p}$. Here, $n$ (\textit{e.g.} $S_0, S_1$) is the electronic state index and $p$ is the photon occupation number, either 0 or 1 in the present work. We consider a cavity photon of frequency 2.8 eV. Therefore, states with $p\ge 2$ lay at least 5.6 eV higher in energy than the ground state, \textit{i.e.} more than 1 eV above our excitation window, which reaches up to 4.5 eV. Due to such high energy difference, they cannot be populated during the dynamics and therefore they are disregarded in our simulations of the photoisomerization dynamics (see Supplemental Note 2). To clearly distinguish the uncoupled states in strong coupling and the electronic states in the zero coupling frameworks, we refer to the set of uncoupled states $\left\{\ket{n,p}\right\}$
with the ket notation, \textit{e.g.} $\ket{S_0,1}$ or $\ket{S_1,0}$, whereas
the zero coupling electronic states $\left\{n\right\}$ are named by the state
label only, \textit{e.g.} $S_0$, $S_1$. The polaritonic eigenstates of
$\hat{H}_{tot}$, labelled as $\ket{R_k}$, are expressed in the $\ket{n,p}$
basis: \\

\eqn{ \ket{R_k}=\sum_{n,p}D^k_{n,p}\ket{n,p}.  } 

The coefficients $D^{k}_{n,p}$ of the uncoupled states in the wavefunction
provide a simple interpretation for the system under strong coupling. The
states with $p=0$ represent no free photon in the cavity, the states with $p=1$ represent one free photon in the cavity and so on. In turn, the time-dependent polaritonic wavefunction can be expressed in either the polaritonic or the uncoupled basis set: 

\eqn{ 
\ket{\Psi(t)}=\sum_{k}C_k(t)\ket{R_k}=\sum_k C_k(t)\sum_{n,p}D^k_{n,p}\ket{n,p}.\label{eq:polexp}}

By the inclusion of the light-molecule interaction, a polaritonic avoided
crossing or conical intersection is originated where the uncoupled states would cross. In Figure \ref{fig:mol_comp}, we show such crossings along the two reactive coordinates: the torsion of the CNNC dihedral and the symNNC respectively. Here, the states labeled as $p>1$ are included in the PESs calculations, yet they are not included in the dynamics presented in the next section.\\

The Rabi splitting between the polaritonic states is proportional to the transition dipole moment between the electronic states at the correspondent crossing geometry for the uncoupled states through eq. \ref{eq:inter}. The magnitude of such splitting represents the coherent energy exchange rate between light and molecule in a confined system. In Figure \ref{fig:mol_comp}c we focus on the polaritonic avoided crossings laying in the \textit{trans} region (CNNC 165$^\circ$). We anticipate that such crossings deeply impacts the photoisomerization mechanism of azobenzene, leading to enhanced \iso{trans}{cis} photoisomerization quantum yield.\\

\section*{Photochemistry on polaritonic states: tuning the photostationary equilibrium} 

In photoreversible processes, the ratio between the quantum yields of the direct and backward process determines the product yield $Q$ at the photostationary state\citep{vetrakova:eps}, as shown in Eq. \ref{eq:photostat}\\

\eqn{
Q=\frac{[c]_\infty}{[c]_\infty+[t]_\infty}=\frac{J_{t\rightarrow c}}{J_{c\rightarrow t}+J_{t\rightarrow c}}=\frac{\varepsilon_{t}\Phi_{t\rightarrow c}}{\varepsilon_{c}\Phi_{c\rightarrow t}+\varepsilon_{t}\Phi_{t\rightarrow c}}
\label{eq:photostat}
}

where $t$ and $c$ respectively refer to the \textit{trans} and \textit{cis} isomers, $J$ is the reaction rate, $\varepsilon$ is the molar extinction coefficient integrated over the excitation wavelength window and $\Phi$ is the quantum yield. The quantities $[c]_{\infty}$, $[t]_\infty$ are the asymptotic concentrations of the \textit{cis} and \textit{trans} isomers respectively, that in this framework correspond to the \textit{cis} and \textit{trans} populations at the end of the dynamics. The ratio between the molar extinction coefficients depends on the excitation wavelength and we shall assume $\varepsilon_t/\varepsilon_c=7.9$ as determined by their integral average over the present excitation interval from the experimental data of azobenzene in methanol\citep{vetrakova:eps}. Such ratio impacts the position of the photostationary state, allowing to shift it selectively towards the \textit{cis} and \textit{trans} isomer depending on the irradiation wavelength. Nevertheless, the tunability is limited by the quantum yields of the individual processes, according to eq. \ref{eq:photostat}. Aiming to manipulate the photostationary state position in azobenzene photoisomerization, we focus on improving the quantum yield of the unfavoured process, namely the \iso{trans}{cis} photoisomerization.\\

To perform the polaritonic photoisomerization simulations, we exploit an on-the-fly surface hopping approach\citep{fregoni:manipulating,persico:direct,persico:decoherence,persico:interplay} and take into account all the nuclear degrees of freedom of azobenzene. Within this framework, the nuclear wavepacket moving on the polaritonic PESs is mimicked by a swarm of independent classical nuclear trajectories (see Methods).\\ 

To build the polaritonic states, we sought a field frequency to maximize the
quantum yields for the $\pi-\pi^*$ \iso{trans}{cis} photoisomerization. We
set the cavity resonant frequency to 2.80 eV, which allows modifying the
crucial region of the first excited state at CNNC close to 180$^\circ$ (detailed in Figure \ref{fig:mol_comp}c) and the surrounding geometries, \textit{i.e.} the region of the PESs where the geometry of the molecule starts to partially twist but it is essentially \textit{trans}. The coupling strength $E_{1ph}$ is 0.002 au, corresponding to a splitting of
$\sim$100 meV with a transition dipole of $\sim$1 a.u for the present case, consistent with the observed 80-100 meV in the experiment by Baumberg and coworkers.\citep{baumberg:singlemol} We sample the ground state distribution at thermostated\citep{bussi:thermo} room temperature. For each sampled configuration, we mimic the excitation by near-UV light, with central wavelength of 313 nm (3.96 eV) and a full bandwidth of 1 eV. The excitation window is chosen to include the absorption spectral features
corresponding to the first $\pi\rightarrow\pi^*$ transitions of \textit{trans}- and
\textit{cis}-azobenzene, though a narrower excitation bandwidth centered at the same frequency yields the same results (see Supplemental Note 1 and Figure \ref{fig:pol_dyn}). Upon the absorption, the
trajectories are vertically excited from the ground state to the polaritonic
states. The excitation procedure is described in the Methods
section\citep{persico:overview}. The polaritonic states initially populated are $\ket{R_3}$, $\ket{R_4}$ and $\ket{R_5}$ which correspond essentially to $\ket{S_2,0}$, $\ket{S_3,0}$ and $\ket{S_4,0}$ in the Franck Condon region. Their populations at time t=0 are 0.76 and 0.21 and 0.03 respectively. In the zero coupling case, the initial populations of the corresponding $S_2$, $S_3$ are 0.78 and 0.22, while $S_4$ is empty.
\\

The polaritonic non-adiabatic dynamics simulations results are reported in
Figure \ref{fig:pol_dyn} (see Supplemental Movie 1
and 2 for the dynamics with strong and zero coupling along the reactive coordinates). The
capability of strong coupling to affect photochemistry is strikingly evident in
Figure \ref{fig:pol_dyn}a, where we compare the population of \textit{trans} and \textit{cis} isomers for the \iso{trans}{cis} photoisomerization process obtained by the zero and strong coupling. Such
populations are evaluated at each time step by counting the number of
trajectories with a CNNC dihedral greater (\textit{trans}) and smaller (\textit{cis}) than 90$^\circ$. The populations are then normalized to the total number of trajectories.\\

\begin{figure*}[ht!]{\includegraphics[width=0.8\linewidth]{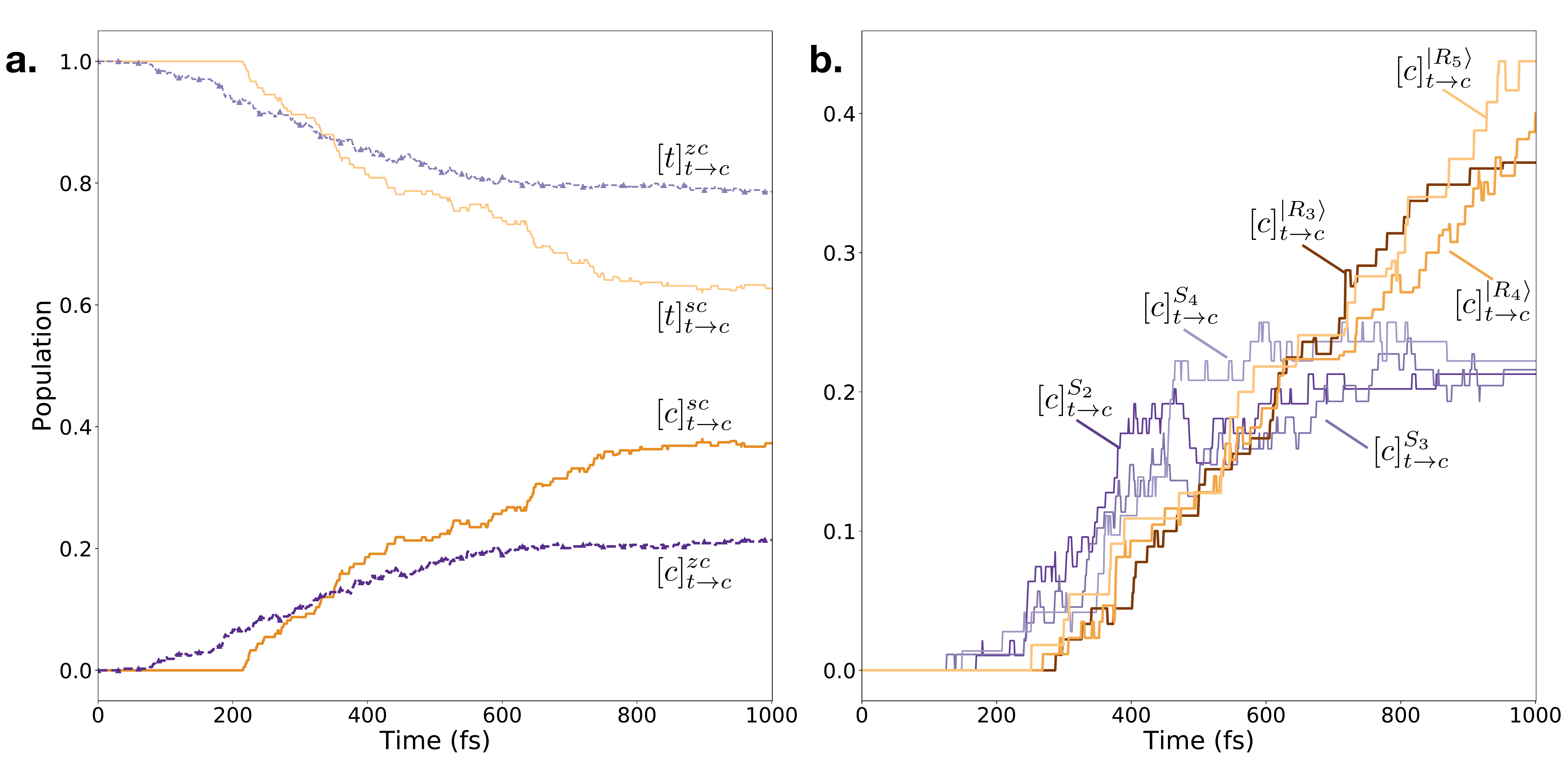}}
	\caption{\textbf{Product-enriched \iso{trans}{cis} photoisomerization of azobenzene under strong coupling } \textbf{a)} Populations of azobenzene \textit{trans} (light) and \textit{cis} (dark) isomers in the zero coupling (purple) and strong coupling (orange) cases for the \iso{trans}{cis} $\pi-\pi^*$ photoisomerization, computed with a photon energy $E_{ph}$ of 2.8 eV and a coupling strength $E_{1ph}$ equal to 0.002 au. \textbf{b)} Comparison between the \textit{cis} formation for processes starting on different electronic/polaritonic states in zero coupling (purple-blue) and strong coupling (orange-red). The individual processes are investigated by running $\sim$100 trajectories. For each pair of initial states in zero coupling and strong coupling, the same sampling is used, \textit{i.e.} $\ket{R_3}$ with $\ket{S_2}$, $\ket{R_4}$ with $S_3$, $\ket{R_5}$ with $S_4$.}\label{fig:pol_dyn} 
\end{figure*}

\begin{figure*}[ht!]{\includegraphics[width=0.8\linewidth]{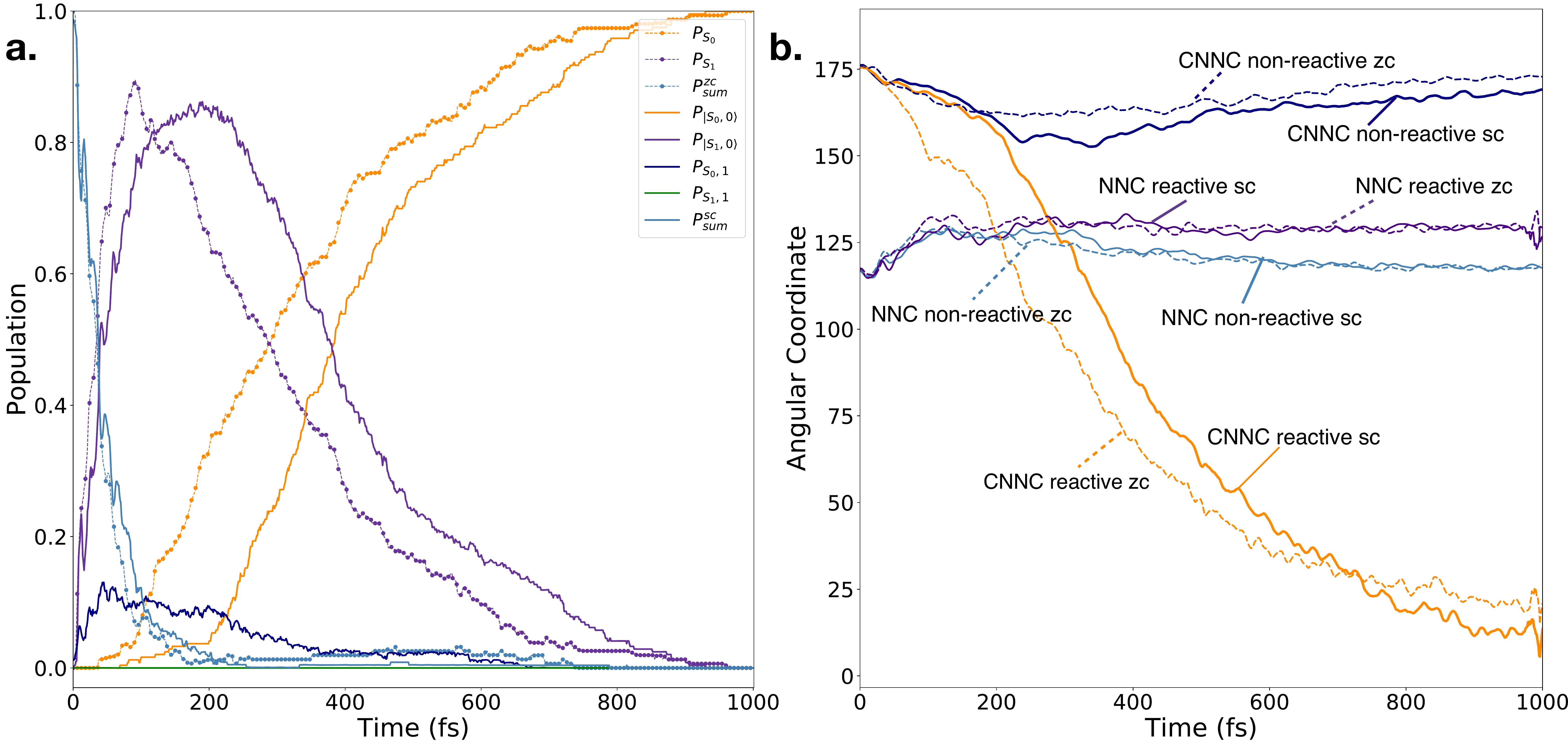}}
	\caption{\textbf{Population and geometrical relaxation dynamics upon photoisomerization} \textbf{a)} Population evolution on the uncoupled states in strong coupling (full lines), directly compared to the zero coupling population evolution involving the same states (dashed lines with markers). The strong coupling population evolution is slowed by the presence of $\ket{S_0,1}$ (blue full line), which is transiently populated during the dynamics. \textbf{b)} CNNC and NNC angles averaged over the reactive and non-reactive trajectories in zero coupling (dashed lines) and strong coupling (full lines), computed as a function of time. See Supplemental Note 4 for the corresponding \iso{cis}{trans} plot.}\label{fig:pol_dyn_2}
\end{figure*}

Remarkably, the \textit{cis} formation is significantly
more efficient for the strong coupling. This is one of the main results of the
present work, as the enhancement of a realistic reaction via electronic strong
coupling has not been reported so far. As a first step to analyze the mechanism
driving such increased yield of product, in Figure \ref{fig:pol_dyn}b we plot the fraction of reactive trajectories (reaching $CNNC<90^\circ$) for each starting state separately. Each of such individual
processes in strong coupling (orange lines) is indeed more efficient than the
corresponding one in zero coupling (purple lines). The strong
coupling processes are on the average slower compared to the zero coupling
case, \textit{i.e.} the torsion around the N=N double bond is delayed, together with the decay to the ground state (Figure \ref{fig:pol_dyn_2}a and \ref{fig:pol_dyn_2}b). Although paradoxically contrasting with the higher yields observed with respect to the zero coupling case, the slower dynamics offers a first hint to explain the change in the mechanism brought about by the strong coupling regime, as detailed later in this work. (See Supplemental Movie 3 for an example of the dynamics along a reactive trajectory).\\ 

The factor capable of both slowing the kinetics and increasing the quantum yields in polaritonic processes is the existence of the $\ket{S_0,1}$ state and its coupling with $\ket{S_1,0}$. Aiming to characterize the nature of the polaritonic states involved in the dynamics and to obtain a more meaningful comparison with the zero-coupling case, it is
convenient to investigate the processes on the uncoupled state basis. To this aim, Figure \ref{fig:pol_dyn_2}a compares the uncoupled states populations (full
lines) with those of the corresponding states in the zero coupling simulation (dashed lines, circle markers). Here, the population of the $\ket{S_2,0}$, $\ket{S_3,0}$ and $\ket{S_4,0}$ manifold is represented as $P_{sum}$ to highlight the relevant processes. The first striking difference is that the $S_1$ state in the zero coupling case is populated quicker than in the strong coupling case. In addition, a longer permanence of the trajectories on the $\ket{S_1,0}$ state is observed in strong coupling, mainly because part of the population oscillates between $\ket{S_1,0}$ and $\ket{S_0,1}$ (see Table S1). Consequently, $\ket{S_1,0}$ (strong coupling) can be found still populated at times where $S_1$ (zero coupling) is already decayed (see Figure \ref{fig:pol_dyn_2}a). The role of the $\ket{S_0,1}$ state in delaying the depletion of $\ket{S_1,0}$ is to act as a supplementary reservoir for the $\ket{S_1,0}$ population during the first 400 fs. In fact, non-radiative electronic state decays from $\ket{S_0,1}$ are blocked since the molecule is in its ground state.\\

The shape of the $\ket{R_1}$ and $\ket{R_2}$ PESs (see Figure \ref{fig:mol_comp}) in the transoid region explains why the torsion is initially delayed in the strong coupling case. Most of the hops that populate these two states go from $\ket{R_3}$ to $\ket{R_2}$ (i.e. essentially $\ket{S_2,0}\rightarrow\ket{S_1,0}$). Subsequently, more transitions back and forth between $\ket{R_1}$ and $\ket{R_2}$ occur, due to the avoided crossing involving $\ket{S_1,0}$ and $\ket{S_0,1}$ (see Table S1). The upper surface, belonging to $\ket{R_2}$, is less favourable than that of $\ket{R_1}$ to the torsional and the symNNC motions (see Figure \ref{fig:mol_comp}a), \textit{i.e.} to the decrease of the CNNC dihedral and to the increase of both NNC angles. By partially populating $\ket{R_2}$, the progress along the reaction coordinate CNNC and the symNNC vibrational excitation are both hindered.\\

The association of slower torsional motion and slower $\ket{S_1,0}$ decay with
higher $\Phi_{t\rightarrow c}$ quantum yield, which
characterizes the strong coupling with respect to the zero
coupling case, is not so intuitive. Still, this effect is reminiscent
of the same joint trends observed in simulations of the
\iso{trans}{cis} photoisomerization in solvents of increasing
viscosity, in agreement with experimental quantum yields and fluorescence lifetimes for the field-free case\citep{cusati:photodynamics}.
A similar hindrance of the motion along the reaction coordinate, caused by strong coupling, was highlighted by Galego \textit{et al}.\citep{galego:suppressing} by full quantum simulations, but unavoidably led to suppression of the photoisomerization because the one-dimensional model cannot account for the competition between radiationless electronic transitions and geometry relaxation. Using a different one-dimensional model, Herrera and Spano showed how strong coupling can instead increase the electron transfer rate in disordered molecular ensembles\citep{spano:coherent}.
\\ 

The reason why a slower progress along the reaction coordinate leads to a
higher quantum yield for the realistic model we are using here can be found in
the shape of the $S_1$, $S_0$ crossing seam. Note that, after leaving the
surroundings of the Franck-Condon region by twisting the N=N bond and/or
increasing the symNNC angles, $\ket{R_1}$ becomes almost pure $\ket{S_1,0}$. In
the new region, its energy gets closer to that of $\ket{R_0}$: a crossing seam
between the two PESs exists. Even more, the crossing seam is practically
unaltered with respect to the zero-coupling case (see Supplemental Note 4 of
the present work and Figure 1 in ref.\citep{cusati:photodynamics}). 
The energy minimum of such seam (optimized conical intersection, CoIn) is found at
a twisted geometry (CNNC=95$^\circ$), the seam is also accessible and coincides with the global minimum in $S_1$, therefore it is accessible even in the absence of vibrational excitation. However, the crossing seam can also be approached at larger
CNNC values by opening the symNNC bond angles, as indicated by our semiempirical PESs and confirmed by accurate ab initio calculations\citep{casellas:seam1,cerullo:azo}. At planar transoid geometries the seam is slightly higher in energy than the Franck-Condon point and much higher than the $S_1$ minimum, so a strong excitation of the symNNC mode is needed to reach it. Recent work based on time-resolved spectroscopy has demonstrated the importance of the symNNC vibration, especially in the case of the $S_0\rightarrow S_2$ excitation\cite{cerullo:azo}.\\

In zero coupling, the symNNC
bending mode is excited once the $S_1$ state is populated by internal
conversion from $S_2$, explainable by comparing the equilibrium
values of the NNC angles in $S_1$ and in $S_0$/$S_2$ (132$^\circ$ versus
118$^\circ$ and 110$^\circ$ at planar geometries). This excitation results in
the opening of the symNNC angle and, in turn, promotes the internal
conversion of $S_1$ to the ground state by making the seam accessible at
transoid regions, resulting in a rather low \iso{trans}{cis}
photoisomerization quantum yield. On the contrary, in strong coupling, the hindering of the
twisting and bending motions discussed above decreases the extent of symNNC
excitation. In fact, with more time spent at transoid geometries, symNNC is
also quenched by vibrational energy transfer to other internal modes and to the
medium. As such, the detrimental effect of the symNNC on the \iso{trans}{cis} 
photoisomerization quantum yield is partially suppressed. The essential role played by (at least) one additional vibrational mode other than the reaction coordinate shows the limitations of one-dimensional models, that may capture some essential features of the dynamics \cite{galego:suppressing} but fail to faithfully describe molecules of useful complexity. Such limitation becomes critical in strong coupling as the PESs and the wavepacket motion are altered by the coupling along all modes. The present case gives a clear example of the need to resort to multi-dimensional models: the trajectories are steered away from the highly-excited symNNC bending (zero coupling) towards the less excited symNNC bending (strong coupling). The alternative pathway due to strong coupling along a secondary coordinate is mainly reflected in the motion along the main isomerization coordinate, resulting in a higher yield pathway not predictable through the one-dimensional models. \\

\begin{figure*}[ht!]{\includegraphics[width=0.8\linewidth]{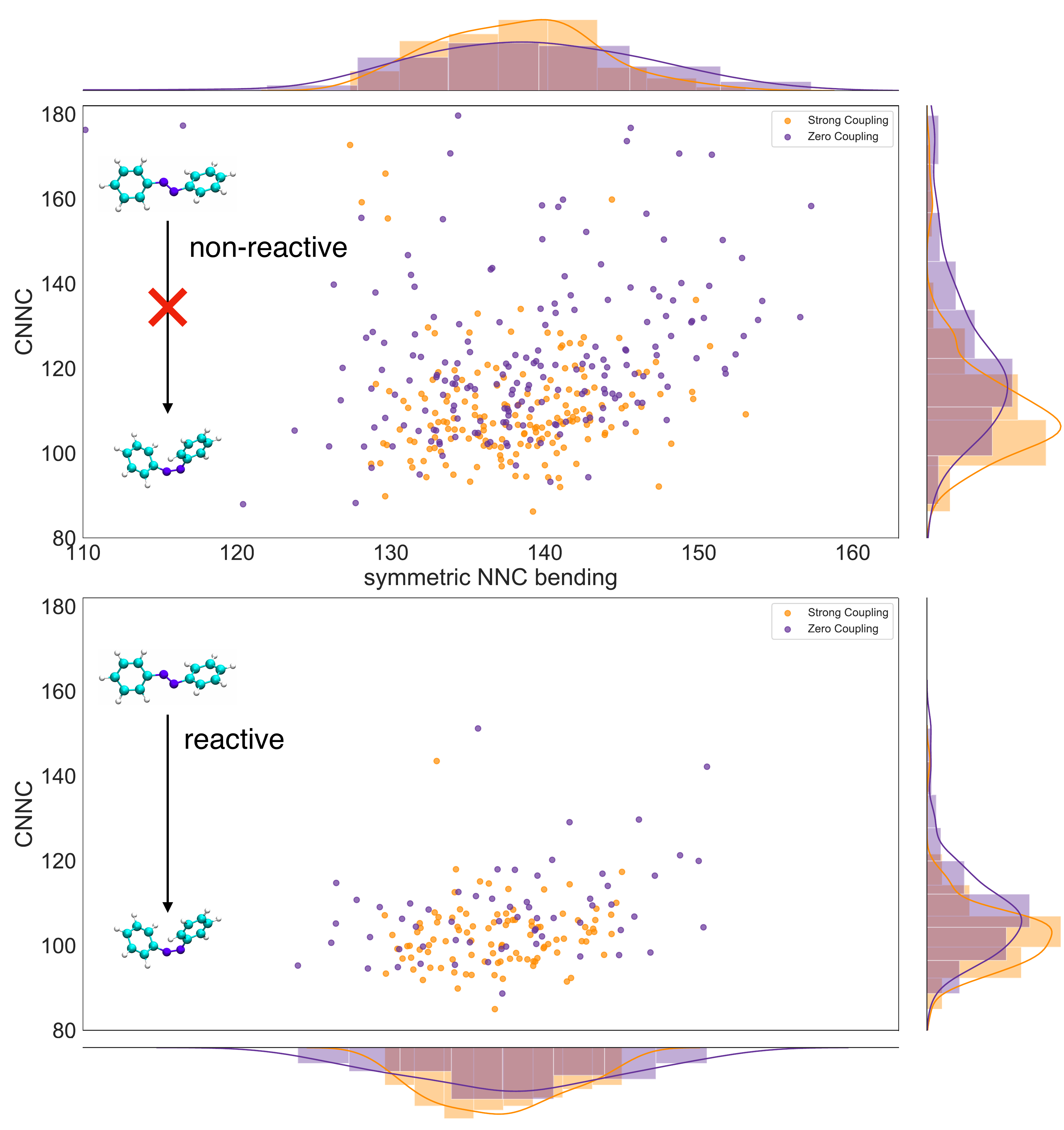}}
	\caption{\textbf{Non-reactive/Reactive photoisomerization dynamics for strong coupling and zero-field upon $S_1-S_0$ hopping} Non-reactive (upper) and reactive (lower) distributions of the reactive coordinates (symNNC, CNNC), computed upon the $S_1-S_0$ hops for the zero-field case (purple) and $\ket{R_1}-\ket{R_0}$ for the strong coupling case (orange). The distributions, in particular the non-reactive one, show that high excitation of symNNC causes hops at less twisted CNNC values, resulting in a lower photoisomerization yield in zero coupling with respect to strong coupling.}\label{fig:geomdyn}
\end{figure*}

The behaviour hereby described is well highlighted in Figure \ref{fig:geomdyn}, where we compare the distribution of the geometrical
coordinates at the moment of the $S_1$-$S_0$ ($\ket{R_1}-\ket{R_0}$) hoppings in
zero coupling (strong coupling), depicted for the non-reactive and
reactive trajectories in the upper and lower panels respectively. Additional data, including the hopping
times, are also provided in Table S1.

The reactive trajectories are shown to hop at CNNC closer to 90$^\circ$,
while the non-reactive ones count many hops at large values of both CNNC and
symNNC. Moreover, a significantly wider distribution of symNNC is observed for
the zero coupling case (purple), a signature that the symNNC is more excited in
zero coupling than in strong coupling. Large symNNC (symNNC$>150^\circ$) in
zero coupling are accompanied by many hops at CNNC$>130^\circ$, confirming that
the excitation of the symmetric NNC vibration promotes the internal conversion at
transoid geometries. The narrower interval of symNNC for the strong coupling
case, instead, causes the trajectories to hop (on the average) at more twisted
geometries, accompanied by a higher probability of successful photoconversion
to the \textit{cis} isomer.\\

Until now, we have shown that the coherent exchange of energy
between light and matter impacts both the kinetics of the
dynamics and the mechanism, resulting in a non-trivial trend in
the quantum yields. To verify the consequence of this result on
photostationary \textit{cis}/\textit{trans} populations, the \iso{cis}{trans}
photoreaction at the same excitation frequency must be simulated
as well. We found that such process in strong coupling shows the same yield with respect to the zero coupling case, $\Phi_{c\rightarrow t}=58\%$ and $\Phi_{c\rightarrow t}=61\%$ respectively. This is consequent to the more favourable slope of the PESs in the \textit{cis} side, which also makes the \iso{cis}{trans} photoisomerization quantum yield insensitive to environmental hindrances\citep{cusati:photodynamics,cantatore:yields,benassi:gold}. Going from the \textit{cis} to the \textit{trans} isomer, such steep PESs make the effect of the $\ket{S_0,1}$ state in the dynamics almost irrelevant, resulting in the \iso{cis}{trans} photoisomerization occurring on much shorter timescales (~150 fs, see Supplemental Note 4) than in the \iso{trans}{cis}. Therefore, the substantial rise of the yields in the \iso{trans}{cis} process is sufficient to push the photostationary state towards the \textit{cis}
isomer.\\ 

When the
system is in its free-photon state $\ket{n,1}$, a loss of the
photon can occur (\textit{e.g.} by leakage from the cavity or absorbed by the cavity walls). As a consequence, the coherent exchange
between light and matter is disrupted and the molecule collapses from
a mixture of $\ket{n',0}$ and $\ket{n,1}$ states to $\ket{n,0}$ only (see
Methods).\\

\begin{figure*}[ht!]{\includegraphics[width=0.8\linewidth]{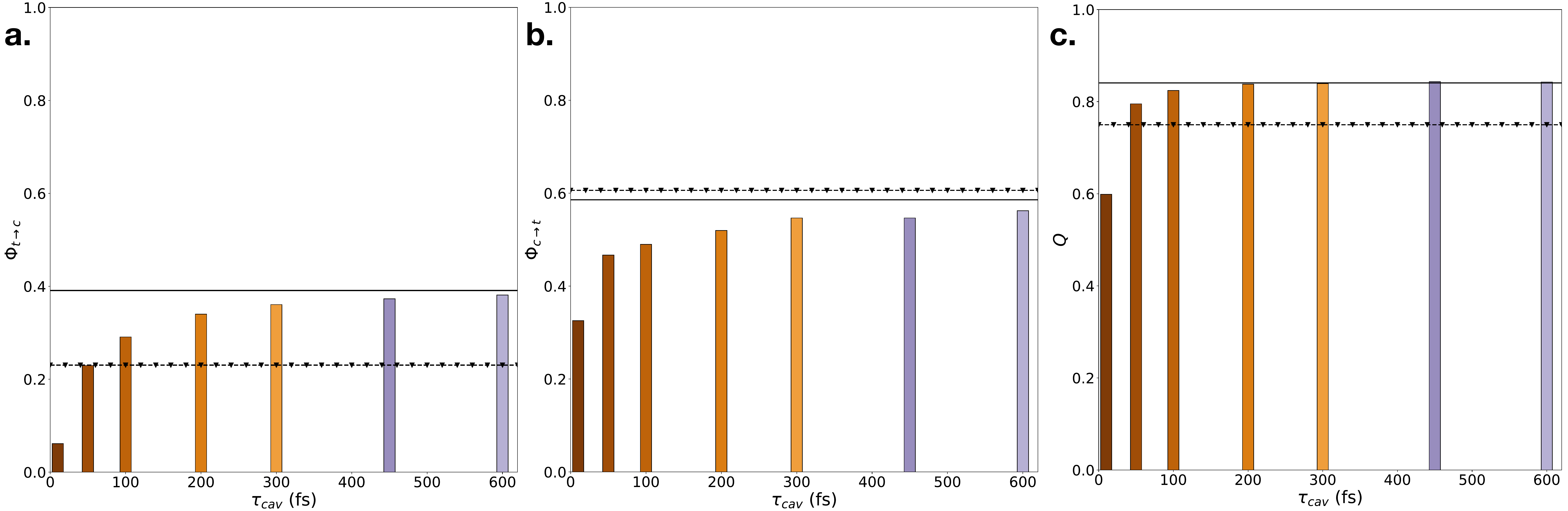}}
	\caption{\textbf{Effect of cavity losses on the photostationary state} Quantum yields comparison for the \textbf{a)} \iso{trans}{cis} and \textbf{b)} \iso{cis}{trans} isomerization in strong coupling as a function of the cavity photon lifetime. The black dotted line is the zero coupling limit, while the full line is the lossless cavity limit. The transient role of $\ket{S_0,1}$ is reflected by lower quantum yields for very lossy cavities with respect to the zero coupling case. \textbf{c)} The product yield at the photostationary state computed by taking into account the forward and backward reaction. The molar extinction coefficients are obtained by their integral average over the present excitation interval, starting from the experimental data reported by Vetrakova and collaborators\citep{vetrakova:eps}. The product yield $Q$ at the photostationary state is shifted towards the strong coupling limit for $\tau_{cav}$ greater than 50 fs.}\label{fig:photostat} 
\end{figure*}

To test how robust the results seen above are with respect to photonic losses
in the resonant cavity, we simulate the \iso{trans}{cis} and \iso{cis}{trans} photoisomerization processes in presence of a finite cavity lifetime $\tau_{cav}$ and compare the so-obtained quantum yields to the zero coupling case (see Figure \ref{fig:photostat}). The photostationary state yield of \textit{cis} product exceeds the zero coupling one for $\tau_{cav}\ge$ 50 fs (Figure
\ref{fig:photostat}a). Remarkably, this time is much shorter than the typical photoisomerization timescale, while intuitively one would expect that cavity lifetimes comparable to the photoisomerization time are needed to observe enhanced reactions. The photoisomerization timescales are longer than the permanence
time of the trajectories on the $\ket{S_0,1}$, which is the only $\ket{n,1}$ state with a non-negligible population at any time. While the decay to the ground
state and the photoisomerization take around 800 fs to be completed,
the average permanence time in $\ket{S_0,1}$ can be estimated to about 35 fs from its time-dependent population. We see then why a photonic loss timescale much
shorter than the timescale of the whole photochemical process is compatible with the observation of strong coupling effects. Below $\tau_\text{cav}$ = 100 fs, however, the \iso{trans}{cis} conversion yield is quite sensitive to cavity lifetime. On the other hand, the \iso{cis}{trans} process is less affected due to the more favourable slope of te PESs and the faster photoisomerization dynamics (see Figure \ref{fig:mol_comp} and Supplemental Note 4).

\subsection*{Discussion/Conclusions}

By building the polaritonic states of azobenzene, we have shown how the molecular complexity can be taken into account for a single molecule strongly coupled to a resonator. The inclusion of a detailed treatment for the molecule and its environment allowed us to investigate the shape of single-molecule polaritons when a manifold of excited states is involved in the strong coupling.\\

We have shown 
that strong coupling deeply affects the dynamical processes taking place on
polaritonic PESs. In particular, we have found a
remarkable increase of the quantum yield for the $\pi-\pi^*$ \iso{trans}{cis} photoisomerization, due to subtle changes in the mechanism: the shape of the polaritonic PESs and the time spent in the one-photon states bring about a lower degree of excitation of the symmetric NNC bending vibration, which is the main cause of early decay from the $S_1$ state in zero-coupling conditions. As a result, under strong coupling more molecules reach a torsion of the N=N bond closer to \textit{cis} before relaxing to the ground state and thus photoisomerize with a higher probability. By taking into account the backward reaction (\iso{cis}{trans}), such effect results in an increase of the photostationary concentration of the \textit{cis} isomer. \\

Through the simulation of a realistic system, \textit{i.e.} by including the effects of environment and cavity losses, we could estimate a minimum cavity lifetime of 50 fs to observe a shift of the photostationary equilibrium towards higher \iso{trans}{cis} photoconversions. Although currently the lifetimes of typical plasmonic nanocavities do not exceed the 10 fs, new experiments are actively devising prototypical setups to achieve high reproducibility\citep{baumberg:singlemol,baumberg:dnaorigami,baumberg:qed} and longer lifetimes for these systems\citep{wang:singlemicrocav,sando:coherent} at the single molecule level. The quickly growing interest in polaritonic applications bodes well for polaritonic devices to be exploited in real-life polaritonic chemistry.\\ 

Our results show promising possibilities in this field. Among them, the enhancement of the quantum yields and photostationary concentrations in experimentally achievable systems opens up a pathway towards a real
control of photochemical reactions (\textit{i.e.} quenching and enhancement). Concerning the role of polaritons in the photochemistry of single molecules, we think that the physics of polariton-mediated reactivity is far from being thoroughly investigated. Among the yet-to-explore possibilities we mention multistate and bielectronic polaritonic processes, such as photoreactions mediated by excitation transfer.\\

\subsection*{Methods}

\paragraph*{Strong coupling Hamiltonian}
The Hamiltonian describing the system is given in eq. \ref{eq:hamiltonian}.
Aiming to include all the degrees of freedom of azobenzene, we exploit a semiempirical AM1 Hamiltonian reparametrized for the first few electronic excited states of azobenzene\citep{cusati:semiemp}. In addition, it includes the molecular interaction with the environment (see next section). The basis on which we build the polaritonic states is the set of electronic-adiabatic singlets $\left\{n\right\}$, from $S_0$ to $S_4$. The cavity Hamiltonian of the quantized electromagnetic field is:
\eqn{
	\hat H_{cav}=\hbar \omega_{cav}\left(\hat b^\dagger \hat b+\frac{1}{2} \right)
}
where $\omega_{cav}$ is the resonator frequency and $\hat b^\dagger$, $\hat b$ are the bosonic creation and annihilation operators. As reported in the main text, the eigenvectors of the non-interacting Hamiltonian $\hat H_{mol}+ \hat H_{cav}$ constitute the uncoupled state basis $\left\{\ket{n,p} \right\}$.
To obtain the polaritonic states (eq. \ref{eq:polexp}) and energies we select a subset of states $\ket{n,p}$ of interest, in which we perfom a CI calculation including the dipolar light-molecule interaction at QM level (eq. \ref{eq:inter}), working in the Coulomb gauge and long wavelength approximation.
The stability of the dipolar approximation has been proven to break up when reaching high couplings \citep{rabl:gauge1,rabl:gauge2,stokes:gauge,distefano:gauge}. To prove the robustness of such approximation in the current case, test calculations have been performed as in the previous work\citep{fregoni:manipulating} (see Supplemental Note 3).

\paragraph*{Inclusion of the environment}

The environment is included at QM/MM level interfaced with the electronic semiempirical Hamiltonian. The molecular Hamiltonian for the system is partitioned as\citep{warshel:qmmm}:
\eqn{\hat {H}_{{mol}}=\hat{H}_{{QM}}+\hat{H}_{{QM/MM}}+\hat{H}_{{MM}}.} The QM part is composed by the azobenzene molecule, the MM part is composed by the cucurbit-7-uril molecule (150 atoms), 710 water molecules and eight frozen layers of gold encapsulating the system (418 atoms each, only van der Waals interactions). The force field used to evaluate the MM part is OPLS-AA contained in the TINKER code\citep{tinker:code}. The QM/MM interactions are modelled by electrostatic embedding plus Lennard-Jones atom-atom potentials\citep{benassi:gold,toniolo:qmmm2,ciminelli:restricted} (See Supplemental Note 1). 

\paragraph*{Surface Hopping on polaritonic states}

After building the molecule embedded in environment and optimizing the geometry at MM level, the starting wavepacket is sampled on the molecular ground state by a QM/MM dynamics. At the end of such dynamics, few hundreds of initial conditions (nuclear phase space point and polaritonic/electronic state) are extracted by evaluating the transition probability from the ground state to the $S_2$,$S_3$,$S_4$ electronic states (zero coupling) or $\ket{R_3},\ket{R_4},\ket{R_5}$ polaritonic states (strong coupling). Both the zero coupling and strong coupling states are excited within the same energy window, \textit{i.e.} centered at 3.96 eV (from 3.46 eV to 4.46 eV). More details can be found in Supplemental Note 1.\\

The non-adiabatic molecular dynamics is perfomed by exploiting the Direct Trajectory Surface Hopping approach\citep{persico:direct}. Few hundreds of classical nuclear trajectories (230 to 270) are computed on-the-fly on the polaritonic PESs independently. The hopping probability between the states is a modified version of Tully's Fewest Switches\citep{tully:tsh}. The modifications added take into account the strong coupling contributions\citep{fregoni:manipulating} and the decoherence corrections needed to properly describe the decoupling of wavepackets travelling on different states\citep{persico:decoherence}.

As usual in surface hopping, the population of a polaritonic state $\ket{R_m}$ is represented by the fraction of trajectories evolving on $\ket{R_m}$ (called the "current" state) at the given time. Consistently, the population of unmixed states $\ket{n,p}$, shown in Figure \ref{fig:pol_dyn_2}a, are obtained by averaging $\left|D^m_{n,p}\right|^2$ over the full swarm of trajectories, where $\ket{R_m}$ is again the current state.

\paragraph*{Cavity Losses} 
The decay probability to account for cavity losses is evaluated through a stochastic approach. In particular, it is taken to be proportional to the square of the coefficients of the uncoupled states $\ket{n,p}$, with $p>0$ ($p=1$ in the present work), composing the time-dependent polaritonic wavefunction (see equation \ref{eq:polexp}):

\eqn{
	P_{{loss}}=\sum^{{n_{st}}}_{{n,p\ge 1}}\frac{1}{\tau_{{cav}}}\Delta t \left| \sum_m C_m D^m_{{n,p}}\right|^2=\sum^{{n_{st}}}_{n,p\ge 1}P_{\ket{n,p}}.
}

\begin{figure*}[ht!]{\includegraphics[width=0.8\linewidth]{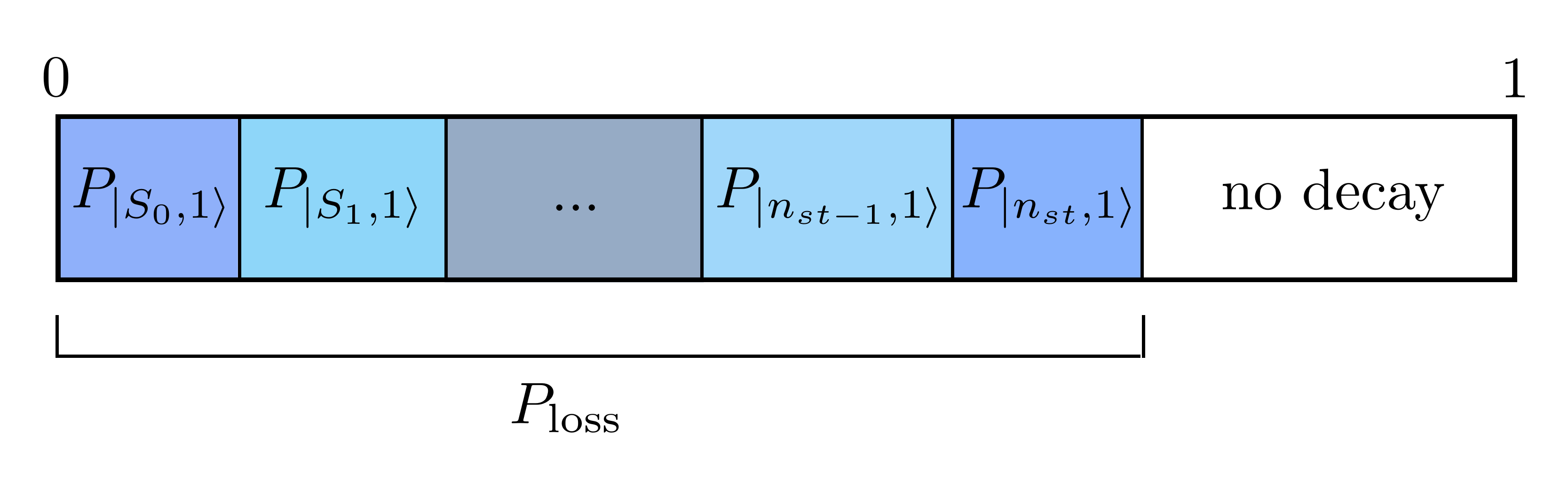}}
	\caption{\textbf{Algorithm to include the photon loss probability} Interval subdivision to evaluate from which state the photon is lost.}\label{fig:decay_sch}
\end{figure*}

Here, $\tau_{\rm{cav}}$ denotes the cavity lifetime while $\Delta t$ is the integration timestep. The decay probabilities referred to each state are indicated as $P_{\ket{n,1}}$
and $n_{st}$ denotes the total number of electronic states included in the
calculation. A uniform random number is generated between 0 and 1 and compared
to the above interval. A check if the random number falls in any sub interval up to $P_{\ket{n_{st},1}}$ is performed. If that is the case, the photon is lost from $\ket{n',1}$. The decay operator $\hat D_{n'}$ is then applied to the polaritonic wavefunction:
\eqn{
	\hat D_{n'}\ket{\Psi}=\ket{n',0}\braket{n',1}{\Psi}.
}
The arrival state $\ket{F}$ is determined by taking the adiabatic state which has the largest overlap $\braket{n',0}{F}$ with the electronic state $\ket{n',0}$.
The dynamics is then resumed by taking $\ket{F}$ as the new current state. We hereby point out that, for our current work, the decay always occurs from the $\ket{S_0,1}$ state, as it is the only state with $p>0$ with a non-negligible population during the dynamics. Even more, the arrival state is always $\ket{R_0}$, as it is almost purely $\ket{S_0,0}$ at all the relevant geometries (see Supplemental Note 2). More generally, the wavefunction after the jump should be written as an electronic wavepacket, mantaining the possible electronic coherence present within the $p>0$ manifold.

\subsection*{Code Availability}

The calculations were based on a locally modified version of MOPAC2002 and TINKER, and are available from G.G. and M.P. upon reasonable request.
The analysis and the supplementary movies are based on \textit{ad-hoc} tools which are available from J.F. upon request.

\subsection*{Acknowledgements}

J.F. and S.C. acknowledge funding from the ERC under the grant ERC-CoG-681285 TAME- Plasmons. G.G. and M.P. acknowledge funding from the University of Pisa, PRA 2017 28 and PRA 2018 36.

\subsection*{Author Contributions}

S.C. initiated this project; J.F., G.G., M.P. and S.C. designed the investigation; J.F. performed the calculations; all the authors contributed to the analysis of results and to the writing of the paper.

\subsection*{Declaration of interests}

The authors declare no competing interests.


\end{document}